\documentclass[%
 reprint,
 superscriptaddress,
nofootinbib,
 amsmath,amssymb,
 aps,
]{revtex4-1}

\usepackage{graphicx}
 \usepackage{dcolumn}
\usepackage{bm}
\usepackage{color}
\usepackage[normalem]{ulem}
\usepackage{hyperref}
\usepackage{braket} 
\usepackage{amsmath}
\usepackage{cancel}

\newcommand{\ccdot}{\!\cdot \!}

\begin{document}
\newcommand{\fref}[1]{Fig.~\ref{fig:#1}} 
\newcommand{\eref}[1]{Eq.~\eqref{eq:#1}} 
\newcommand{\erefn}[1]{ (\ref{eq:#1})}
\newcommand{\erefs}[2]{Eqs.~(\ref{eq:#1}) - (\ref{eq:#2}) } 
\newcommand{\aref}[1]{Appendix~\ref{app:#1}}
\newcommand{\sref}[1]{Section~\ref{sec:#1}}
\newcommand{\cref}[1]{Chapter~\ref{ch:.#1}}
\newcommand{\tref}[1]{Table~\ref{tab:#1}}

\newcommand{\nn}{\nonumber \\}  
\newcommand{\nnl}{\nonumber \\}  
\newcommand{\nl}{& \nonumber \\ &}
\newcommand{\bnl}{\right .  \nonumber \\  \left .}
\newcommand{\dbnl}{\right .\right . & \nonumber \\ & \left .\left .}

\newcommand{\beq}{\begin{equation}} 
\newcommand{\eeq}{\end{equation}} 
\newcommand{\ba}{\begin{array}}  
\newcommand{\ea}{\end{array}} 
\newcommand{\bea}{\begin{eqnarray}}  
\newcommand{\eea}{\end{eqnarray} }  
\newcommand{\be}{\begin{eqnarray}}  
\newcommand{\ee}{\end{eqnarray} }  
\newcommand{\bal}{\begin{align}}
\newcommand{\eal}{\end{align}}   
\newcommand{\bi}{\begin{itemize}}  
\newcommand{\ei}{\end{itemize}}  
\newcommand{\ben}{\begin{enumerate}}  
\newcommand{\een}{\end{enumerate}}  
\newcommand{\bc}{\begin{center}}
\newcommand{\ec}{\end{center}} 
\newcommand{\bt}{\begin{table}}
\newcommand{\et}{\end{table}}  
\newcommand{\btb}{\begin{tabular}}
\newcommand{\etb}{\end{tabular}}  
\newcommand{\bvec}{\left ( \ba{c}}
\newcommand{\evec}{\ea \right )}

\newcommand{\cO}{{\mathcal O}} 
\newcommand{\co}{{\mathcal O}} 
\newcommand{\cL}{{\mathcal L}} 
\newcommand{\cl}{{\mathcal L}} 
\newcommand{\cM}{{\mathcal M}}

\newcommand{\const}{\mathrm{const}}

\newcommand{\ev}{ \mathrm{eV}}
\newcommand{\kev}{\mathrm{keV}}
\newcommand{\mev}{\mathrm{MeV}}
\newcommand{\gev}{\mathrm{GeV}}
\newcommand{\tev}{\mathrm{TeV}}

\newcommand{\mpl}{M_{\mathrm Pl}}

\def\mgut{\, M_{\rm GUT}}
\def\tgut{\, t_{\rm GUT}}
\def\mpl{\, M_{\rm Pl}}
\def\mkk{\, M_{\rm KK}}
\newcommand{\msusy}{M_{\rm soft}}

\newcommand{\dslash}[1]{#1 \! \! \! {\bf /}}
\newcommand{\ddslash}[1]{#1 \! \! \! \!  {\bf /}}

\def\ads{AdS$_5$\,}
\def\adse{AdS$_5$}
\def\intdk{\int {d^4 k \over (2 \pi)^4}} 

\def\ra{\rangle}
\def\la{\langle}  

\def\sgn{{\rm sgn}}
\def\pa{\partial}  
\newcommand{\dlr}{\overleftrightarrow{\partial}}
\newcommand{\Dlr}{\overleftrightarrow{D}}
\newcommand{\re}{{\mathrm{Re}} \,}
\newcommand{\im}{{\mathrm{Im}} \,}
\newcommand{\tr}{\mathrm T \mathrm r}  

\newcommand{\Ra}{\Rightarrow}
\newcommand{\lra}{\leftrightarrow}
\newcommand{\llra}{\longleftrightarrow}

\newcommand\simlt{\stackrel{<}{{}_\sim}}
\newcommand\simgt{\stackrel{>}{{}_\sim}}   
\newcommand{\zt}{$\mathbb Z_2$ }

\newcommand{\ha}{{\hat a}}
\newcommand{\hab}{{\hat b}}
\newcommand{\hac}{{\hat c}} 

\newcommand{\ti}{\tilde}  
\def\hc{{\rm h.c.}} 
\def\ov{\overline}  
  

\newcommand{\eps}{\epsilon}
\newcommand{\eS}{\epsilon_S}
\newcommand{\eT}{\epsilon_T}
\newcommand{\eP}{\epsilon_P}
\newcommand{\eL}{\epsilon_L}
\newcommand{\eR}{\epsilon_R}
\newcommand{\teps}{{\tilde{\epsilon}}}
\newcommand{\teS}{{\tilde{\epsilon}_S}}
\newcommand{\teT}{{\tilde{\epsilon}_T}}
\newcommand{\teP}{{\tilde{\epsilon}_P}}
\newcommand{\teL}{{\tilde{\epsilon}_L}}
\newcommand{\teR}{{\tilde{\epsilon}_R}}
\newcommand{\eLc}{{\epsilon_L^{(c)}}}
\newcommand{\eLv}{{\epsilon_L^{(v)}}}
\newcommand{\eSP}{\epsilon_{S,P}}
\newcommand{\teSP}{{\tilde{\epsilon}_{S,P}}}

\newcommand{\lz}{\lambda_z}
\newcommand{\dgz}{\delta g_{1,z}}
\newcommand{\dkg}{\delta \kappa_\gamma}

\def\cog{\color{OliveGreen}}
\def\cor{\color{Red}}
\def\copu{\color{purple}}
\def\coro{\color{RedOrange}}
\def\coma{\color{Maroon}}
\def\cob{\color{Blue}}
\def\cobr{\color{Brown}}
\def\cobl{\color{Black}}
\def\cost{\color{WildStrawberry}}

\newcommand{\tl}{{\tilde{\lambda}}}
\newcommand{\dll}{{\frac{\delta\lambda}{\lambda}}}

\title{Bases of massless EFTs via momentum twistors}

\author{Adam Falkowski}

\affiliation{Universit\'{e} Paris-Saclay, CNRS/IN2P3, IJCLab, 91405 Orsay, France}

\begin{abstract}

I present a novel method of deriving a basis of contact terms in massless effective field theories (EFTs).
It relies on the parametrization of $N$-body kinematics via the so-called momentum twistors.  
A basis is constructed directly at the amplitude level, without using fields or Lagrangians.  
The method consists in recasting any local contact term as a sum of rational functions built from Lorentz-invariant contractions of momentum twistors. 
The end result is equivalent to constructing a basis of higher-dimensional operators in an EFT Lagrangian, 
however it is considerably simpler, especially for theories with higher-spin particles.  
The method is applied to contact terms in 4-point amplitudes. 
I provide a compact algebraic formula for basis elements for any helicity configuration of the external particles, and I  illustrate its usage with several physically relevant examples.

\end{abstract}

\maketitle

\section{Introduction}
\label{sec:intro}

In a relativistic quantum theory, amplitudes can be calculated using Feynman rules derived from a Lagrangian. 
The alternative is to directly construct on-shell amplitudes using the basic principles of Poincar\'e symmetry, locality, and unitarity, without using the crutches of fields and Lagrangians~\cite{Benincasa:2007xk,Elvang:2013cua,Cheung:2017pzi,Arkani-Hamed:2017jhn}.  
This approach is based on the fact that, on kinematic poles, residues of an $N$-point amplitude  factorize into  products of lower-point  amplitudes. 
For example, a tree-level amplitude with 4 massless particles on the external legs can be represented as 
\bea 
A(1234)  &= &- {A(12 \hat p_s) A(34 p_s) \over s}
 - {A(13 \hat p_t) A(24 p_t) \over t}
 \nnl && 
  - {A(14 \hat p_u) A(23 p_u) \over u}  + C(1234) , 
\eea 
where  $1\dots 4$ label incoming particles, 
$p_s \equiv p_1 + p_2$, $p_t \equiv  p_1 + p_3$, $p_u \equiv  p_1 + p_4$, 
the Mandelstam invariants are $x \equiv p_x^2$  for $x = s,t,u$, 
and the hats denote outgoing particles.
The last piece stands for {\em contact terms}, which are regular functions of $s,t,u$ without poles or other singularities, therefore they are not connected to lower-point amplitudes by unitarity.  
All in all, an on-shell 4-point amplitude can be bootstrapped from 3-point ones, up to contact terms. 
The latter are the focus of this paper. 

One can fix the contact terms through physically motivated assumptions about the high-energy or analytic behavior of the amplitudes.  
This path is relevant for certain important theories, such as QCD or general relativity (GR),
which are completely fixed by their 3-point amplitudes and do not admit any free parameters entering at $N > 3$~\cite{Britto:2005fq,Benincasa:2007qj}.  
EFT takes exactly the  opposite path:
the contact terms are  assumed to have the most general form allowed by the symmetries, 
and are organized as an expansion in inverse powers of a high mass scale $\Lambda$:   
\beq
C(1234) = \sum_{D = 4}^\infty \sum_k {c_{D,k} O_{D,k}(1234)  \over \Lambda^{D-4}},  
\eeq  
where  $O_{D,k}$ are basis elements spanning the space of all possible Lorentz-invariant contact terms.  
Each $O_{D,k}$ is a regular function of Mandelstam variables with mass dimension $[\rm mass]^{D-4}$, 
while $c_{D,k}$ are free parameters called the Wilson coefficients.  
Formally, the contact terms depend on an infinite number of Wilson coefficients. 
However, for processes with a characteristic energy scale $E \ll \Lambda$ only $O_{D,k}$ with low enough $D$ are numerically important, and the amplitude can be well approximated using a finite number of $c_{D,k}$ and $O_{D,k}$.  

There remains the highly non-trivial issue of writing down all possible $O_{D,k}$, given the quantum numbers (in particular spin and helicity) of the external particles.
In the Lagrangian language, the parallel problem is constructing all independent Lorentz-invariant local operators of canonical dimension $D$. 
That task can be systematically organized thanks to the Hilbert series techniques~\cite{Lehman:2015via,Henning:2015daa,Henning:2017fpj}. 
Alternatively, for massless EFTs one can bypass Lagrangians and construct contact terms directly  at the amplitude level~\cite{Cheung:2014dqa,Cheung:2016drk,Shadmi:2018xan} using the spinor helicity variables ({\em helicity spinors}, in short). 
This method may be simpler, especially for higher spins, as spinors encode helicity information in a transparent way. 
Recently, Ref.~\cite{Henning:2019enq} proposed an algorithm for constructing a basis $O_{D,k}$ as harmonic modes of the physical manifold of helicity spinors parametrizing $N$-point amplitudes.
See also Refs.~\cite{Ma:2019gtx,Durieux:2019siw} for other amplitude-based approaches to constructing contact  terms. 

In this paper I propose a novel way of constructing a basis of contact terms in massless EFTs. 
The departure point is the parameterization of N-point amplitudes using the {\em momentum twistor} variables~\cite{Hodges:2009hk,Elvang:2013cua}.
Namely, the kinematic data can be encoded in $N$ spinor pairs $Z_i = (\lambda_i, \tilde \mu_i)$. 
From these, the  four-momenta $p_i$ of all external particles can be reconstructed, and they automatically satisfy the on-shell condition ($p_i^2=0$) and momentum conservation ($\sum_i p_i = 0$).
The basic facts about momentum twistors are summarized in \sref{flash}. 
These are very convenient and natural variables on the space of $N$-body kinematics.   
In particular, they greatly simplify the task of constructing independent Lorentz invariants from the kinematic data.    
\sref{base} shows how to trade local contact terms for rational functions of Lorentz invariant $(\lambda_i \lambda_j)$  and $(\tilde \mu_i \tilde \mu_j)$ spinor contractions.
Among these, a set of independent basis elements is identified, 
from which any contact term can be constructed at a given order in the EFT expansion and for a given helicity configuration of external particles. 
In this paper I focus on 4-point amplitudes, however the method can be generalized to higher $N$.
For $N=4$, the candidate basis elements are characterized by a compact formula in \eref{BASE_candidate}.
At a fixed EFT order they are labeled by one integer bounded to a finite range, 
thus they can be enumerated order by order in the EFT expansion.  
\sref{ex} illustrates this method with a number of simple examples rooted in physically relevant theories.    
A slightly more involved example of constructing contact terms and scattering amplitudes in the EFT extension of GR is relegated to \aref{gr}.

I work in 4 spacetime dimensions with the mostly minus metric: $\eta_{\mu \nu} = {\rm diag}(1,-1,-1,-1)$. 
The Lorentz algebra decomposes into $SU(2) \times SU(2)$, with holomorphic and anti-holomporphic spinors $\lambda_\alpha$ and $\tilde \lambda_{\dot \alpha}$  transforming under the respective $SU(2)$ factors. 
For the spinors I adopt the conventions of  Ref.~\cite{Dreiner:2008tw}. 
Spinor indices are raised with the antisymmetric tensor $\epsilon^{\alpha \beta}$ and  lowered with $\epsilon_{\alpha \beta}$:  $\lambda^\alpha = \epsilon^{\alpha \beta} \lambda_\beta$, $\lambda_\alpha = \eps_{\alpha \beta} \lambda^\beta$, and idem for $\tilde \lambda$,  with the convention $\epsilon^{12} = - \epsilon_{12} = 1$.  
The Lorentz invariant spinor contractions are 
$\lambda_i^{\alpha} \lambda_{j \, \alpha} \equiv (\lambda_i \lambda_j) \equiv \langle ij \rangle$, 
$\tilde \lambda_{i \, \dot \alpha} \tilde \lambda_{j}^{\dot \alpha}\equiv (\tilde \lambda_i \tilde \lambda_j)  \equiv [ ij ]$. 
Vector and spinor Lorentz indices can be traded with the help of the sigma matrices 
$[\sigma^\mu]_{\alpha \dot \beta}  = ({\bf 1}, \vec\sigma )$, where $\vec  \sigma$ are the Pauli matrices. 
I abbreviate $p\ccdot \sigma \equiv p_\mu \sigma^\mu$.

\section{Flash review of momentum twistors}
\label{sec:flash}

I start by reviewing the momentum twistor description of massless $N$-body kinematics in four spacetime dimensions, following closely the presentation in Ref.~\cite{Elvang:2013cua}. 
Consider a set of four-momenta $p_i$, $i = 1\dots N$,  subject to the on-shell conditions 
$p_i^2 = 0$ 
and momentum conservation $\sum_{i=1}^N p_i = 0$. 
The restrictions on $p_i$ may be inconvenient to work with, 
and for many applications it is beneficial to introduce different variables that trivialize the constraints.   
The on-shell conditions are dealt with by introducing the helicity spinors, that is  $N$ holomorphic  and anti-holomorphic  2-component spinors   $\lambda_{i}$,  $\tilde \lambda_{i}$ related to the four-momenta by   
$p_i \ccdot \sigma  = \lambda_{i} \tilde \lambda_{i}$. 
This  trivializes the on-shell constraints, in the sense that an arbitrary pair  $(\lambda_i, \tilde \lambda_i)$  defines a (possibly complex) $p_i$ that automatically satisfies $p_i^2 = 0$. 
As a bonus, the transformation $\lambda_i  \to t^{-1}_i \lambda_i$, $\tilde \lambda_i  \to t_i \tilde \lambda_i$ does not change $p_i$ therefore it represents the little group action on the particle $i$.  
However, momentum conservation is not automatic in these variables, and implies one non-linear constraint on  the $N$ spinors $\lambda_i$ and $\tilde \lambda_i$. 
The idea behind the momentum twistors is to trade the helicity spinors into a different set of variables so as to trivialize the momentum conservation as well.   
To this end, one defines the dual coordinates $y_i$ via the relation 
\beq
\label{eq:MT_defy}
p_i =  y_i - y_{i-1} \quad \Rightarrow \quad y_i = y_N  + \sum_{j=1}^{i} p_j  , 
\eeq 
with the cyclic identification $y_0 = y_N$.
One can think of  $y_i$ as vertices of a polygon in the dual coordinate space, and of $p_i$ as the sides of that polygon.   
Note that $y_i$ have units of $[\rm mass]^1$, unlike the spacetime coordinates. 
The four-momenta $p_i$ do not depend on $y_N$: they do not change when the polygon is moved around in the dual spacetime.   
This is just the translation invariance in the dual coordinate space. 
One can always gauge-fix it,  e.g. by setting $y_N =0$.  

Introducing $y_i$ via \eref{MT_defy} trivializes the momentum conservation. 
However the goal is to construct variables that trivialize both the momentum conservation and the on-shell conditions. 
This is achieved by introducing a set of $N$ anti-holomorphic spinors $\tilde \mu_i$ defined as 
\beq
\label{eq:MT_defmu}
\tilde \mu_i = \lambda_i \sigma \ccdot y_i = \lambda_i \sigma \ccdot y_{i-1}  .  
\eeq  
Given that $\lambda_i$ have dimension $[\rm mass]^{1/2}$, $\tilde \mu_i$ have dimension $[\rm mass]^{3/2}$. 
The spinor pairs $Z_i = (\lambda_i, \tilde \mu_i)$ are called the momentum twistors. 
Any set of $N$ such pairs automatically defines an $N$-body kinematics with $p_i^2=0$ and $\sum_{i=1}^N p_i = 0$.  
In order to see this, note that \eref{MT_defmu} can be solved for the dual coordinate: 
\beq
\label{eq:MT_soly}
y_i \ccdot \sigma = {\lambda_{i+1} \tilde \mu_{i} - \lambda_{i} \tilde \mu_{i+1} \over 
(\lambda_{i}  \lambda_{i+1})}. 
\eeq 
Therefore, starting from $Z_i$ one can reconstruct all $y_i$, and thus  $p_i$ via \eref{MT_defy},  
Furthermore, one can show that these four-momenta can be decomposed as  
$p_i \ccdot \sigma  = \lambda_i \tilde \lambda_i$,  where 
\beq
\label{eq:MT_tlambda}
\tilde \lambda_i  = - {\tilde \mu_{i-1} (\lambda_i \lambda_{i+1}) + \tilde \mu_{i} (\lambda_{i+1} \lambda_{i-1}) 
+ \tilde \mu_{i+1} (\lambda_{i-1} \lambda_{i})
\over (\lambda_{i-1} \lambda_{i}) (\lambda_i \lambda_{i+1}) } . 
\eeq  
From \eref{MT_soly} and \eref{MT_tlambda}  it follows that $\lambda_i$ and $\tilde \mu_i$ transform in the same way  under the little group: $Z_i \to t_i^{-1} Z_i$. 
In other words, $Z_i$ are defined projectively: an independent rescaling of each $Z_i$ does not change $p_i$. 
In the literature, momentum twistors are most often used in superconformal frameworks (see e.g. Ref.~\cite{ArkaniHamed:2009dn}), in which case $Z_i$ are unbreakable building blocks of the amplitudes. 
In this paper however I am interested in more down-to-earth theories. 
I will treat $\lambda_i$ and $\tilde \mu_i$ as separate building blocks, from which I construct Lorentz-invariant  holomorphic $(\lambda_i \lambda_j)$ and anti-holomorphic $(\tilde \mu_i \tilde \mu_j)$ spinor contractions.

Let us compare  the number of degrees of freedom on the four-momentum and momentum twistor sides. 
In complex kinematics, each $p_i$ has 6 real degrees of freedom after imposing the on-shell condition. 
Momentum conservation fixes 8 degrees of freedom, thus  the manifold of $N$-body kinematics is $(6N-8)$-dimensional.  
On the twistor side,  fixing the translation invariance by setting $y_N =0$ corresponds via \eref{MT_defmu} to setting $\tilde \mu_1 = \tilde \mu_N =0$.  
This leaves $N$ spinors $\lambda_i$, and $N-2$ spinors $\tilde \mu_i$. 
Each spinor has 4 real degrees of freedom, however 2 degrees of freedom in  each $\lambda_i$ can be removed by  little group transformations  $Z_i \to t^{-1}_i Z_i$. 
This leaves $2 N + 4 (N-2) = 6N - 8$ degrees of freedom, in agreement with the dimensionality of the kinematics manifold.  
This shows that the projective space spanned by $N$ momentum twistors $Z_i$ is in 1-to-1 correspondence  (after gauge-fixing the translations) with the space of all independent massless $N$-body kinematics.

Instead of $Z_i = (\lambda_i, \tilde \mu_i)$,  one could represent the kinematic data by momentum anti-twistors 
$\tilde Z_i \equiv (\mu_i, \tilde \lambda_i)$ related to the dual coordinates by 
$\mu_i = y_i \ccdot \sigma \tilde \lambda_i  = y_{i-1}\ccdot \sigma \tilde \lambda_i$. 
In the spinor helicity variables, parity exchanges  $\lambda_i \leftrightarrow \tilde \lambda_i$.    
The binary choice of either $Z_i$ or $\tilde Z_i$ to represent the kinematic data is not parity invariant, 
therefore parity is not manifest in momentum twistor variables.   

\section{Contact terms via momentum twistors}
\label{sec:base}

Contact terms are functions of the helicity spinors that can be written in the form
\beq 
\label{eq:contact}
O = \Pi_{i<j}  (\lambda_i \lambda_j)^{n_{ij}} (\tilde \lambda_i \tilde \lambda_j)^{\tilde  n_{ij} },
\quad 
i,j = 1 \dots N , 
 \eeq   
where the exponents $n_{ij}$ and $\tilde n_{ij}$ are non-negative integers. 
The little group dictates that the exponents are related to the helicities of the external particles by 
\beq
\label{eq:BASE_hk}
2 h_k = \sum_{j \neq k } \left ( \tilde n_{jk}  - n_{jk} \right ) . 
\eeq
\eref{contact} is manifestly local, however expressions with different $n_{ij}$ and $\tilde n_{ij}$ may be dependent. 
For example, 
$\langle 12 \rangle^2  \langle 34 \rangle^2$, 
$\langle 13 \rangle^2  \langle 24 \rangle^2$, 
$\langle 14 \rangle^2  \langle 23 \rangle^2$, 
and $\langle 12 \rangle  \langle 34 \rangle \langle 13 \rangle  \langle 24 \rangle$ 
are all legal 4-point contact terms, however only 3 out of these 4 are independent. 
In this section I lay out an algorithm to construct a basis of independent contact terms for $N$-point amplitudes.
To this end, I parametrize $N$-body kinematics by momentum twistors 
$(\lambda_i, \tilde \mu_i)$ subject to the gauge fixing condition $\tilde \mu_1 = \tilde \mu_N = 0$.
Then I recast \eref{contact} as a sum of independent rational functions of 
$(\lambda_i \lambda_j)$ and  $(\tilde \mu_i \tilde \mu_j)$. 
This allows me to identify a basis from which any contact term can be constructed. 
Of course, such a basis is infinite,  however its  elements can be organized according to their mass dimensions,  
which directly translates into EFT expansion in powers of $1/\Lambda$. 
At a given EFT order and for a given helicity configuration the number of  basis elements is finite. 

Below I perform this construction for $N=4$, and then comment on extending it to higher $N$. 
 
\subsection{4-point}
\label{sec:4p}

I start with \eref{contact} for $N=4$. 
The $(\tilde \lambda_i \tilde \lambda_j)$ spinor contractions can be traded for momentum twistors using \eref{MT_tlambda}:  
\bea 
\label{eq:BASE_tildetomt}
& 
[12] = {(\tilde \mu_2 \tilde \mu_3) \over \langle 12 \rangle \langle 2 3 \rangle}, 
\ \ 
[13] = - {(\tilde \mu_2 \tilde \mu_3) \langle 2 4 \rangle \over \langle 12 \rangle \langle 2 3 \rangle \langle 3 4 \rangle }, 
\ \ 
[14] = {(\tilde \mu_2 \tilde \mu_3) \over \langle 12 \rangle \langle 34  \rangle}, 
& \nnl \,  & 
[23] = {(\tilde \mu_2 \tilde \mu_3) \langle 1 4 \rangle \over \langle 12 \rangle \langle 2 3 \rangle \langle 3 4 \rangle }, 
\ \
[24] = - {(\tilde \mu_2 \tilde \mu_3) \langle 1 3 \rangle \over \langle 12 \rangle \langle 2 3 \rangle \langle 3 4 \rangle }, 
\ \
[34] = {(\tilde \mu_2 \tilde \mu_3) \over \langle 23 \rangle \langle 34  \rangle}. 
\nnl
\eea 
The fact that on the momentum twistor side there is only one possible anti-holomorphic contraction  $(\tilde \mu_2 \tilde \mu_3)$ greatly simplifies the proceedings for $N=4$.
Plugging  \eref{BASE_tildetomt} into \eref{contact} yields a representation of any 4-point local term as a rational function of $(\tilde \mu_2 \tilde \mu_3)$ and $(\lambda_{i}\lambda_j)$ contractions:\footnote{%
In this discussion signs and numerical factors are irrelevant, and they are dropped between equations.}
\bea
O & =&    (\tilde \mu_2 \tilde \mu_3)^n 
\langle 12 \rangle^{n_{12} + \tilde n_{34} - n}
\langle 13 \rangle^{n_{13} + \tilde n_{24} }
\langle 14 \rangle^{n_{14} + \tilde n_{23} } 
\nnl & \times & 
\langle 23 \rangle^{n_{23} + \tilde n_{14} - n}
\langle 24  \rangle^{n_{24} + \tilde n_{13}}
\langle 34  \rangle^{n_{34} + \tilde n_{12} - n} , 
\eea 
where $n = \sum_{i<j} \tilde n_{ij}$, which implies $n \geq 0$. 
One spinor contraction, e.g.  $\langle 14 \rangle$, can be eliminated via the Schouten identity
$\langle 12 \rangle \langle 34 \rangle -  \langle 13 \rangle \langle 24 \rangle  + \langle 14 \rangle \langle 23\rangle = 0$. 
Then $O$ is represented as a {\em sum} of rational functions of momentum twistor contractions: 
\bea
\label{eq:BASE_twistorcandidate0}
O & =&   { (\tilde \mu_2 \tilde \mu_3)^n \over  \langle 23 \rangle^n}
\langle 23 \rangle^{n_{23} -n_{14}   + \tilde n_{14} -  \tilde n_{23}}
\langle 12 \rangle^{n_{12} + n_{14}+ \tilde n_{24} + \tilde n_{34} - n - \alpha}
\nnl & \times & 
\langle 13 \rangle^{n_{13} + \tilde n_{24} + \alpha}
\langle 24  \rangle^{n_{24} + \tilde n_{13} +\alpha}
\langle 34  \rangle^{n_{34} + \tilde n_{12} - n} , 
\eea 
and $\alpha$ is an integer in the range $[0,n_{14}+\tilde n_{23}]$. 
In the last step I traded $n_{ij}$ and $\tilde n_{ij}$ for external helicities using \eref{BASE_hk}.  
All in all, any contact term can be expressed as a sum of basis elements:   
\bea
\label{eq:BASE_twistorcandidate}
O_{n,k}^{h_1 h_2h_3h_4} & = & \left ( (\tilde \mu_2 \tilde \mu_3) \over \langle 2 3 \rangle   \right )^n 
\left ( \langle 13 \rangle \langle 24 \rangle  \over  \langle 12 \rangle \langle 34 \rangle   \right )^k 
\langle 23 \rangle^{h_1 - h_2 - h_3 + h_4} 
 \nnl & \times & 
 \langle 12 \rangle^{-2 h_1}     \langle 24 \rangle^{h_1 - h_2 + h_3 - h_4}   \langle 34 \rangle ^{-h_1 + h_2 - h_3 - h_4}  ,  
 \nnl   
\eea
where $k =n_{13} + \tilde n_{24} + \alpha \geq 0$. 
For a given helicity configuration, the candidate basis elements are parametrized by two integers: $n$ and $k$.
The former controls the canonical dimension, which is related to $n$ by  
\beq
\label{eq:BASE_dim}
D = 2 n + 4 - \sum_i h_i .
\eeq 
The latter labels basis elements for each canonical dimension, that is at each EFT order.  
An important point is that, for $n$ fixed,  $k$ is constrained to a finite range for $O_{n,k}^{h_1 h_2h_3h_4} $ to   possibly be a local term. 
This can be seen by looking at the scaling: 
$(\tilde \mu_2 \tilde \mu_3) \sim s \sqrt{u}$,  
  $\langle 23 \rangle \sim \sqrt{u}$, 
$\langle 12 \rangle \sim \sqrt{s}$,  $\langle 34 \rangle \sim \sqrt{s}$,  
$\langle 13 \rangle \sim \sqrt{t}$,  $\langle 24 \rangle \sim \sqrt{t}$. 
A contact term must be non-singular when $s$ or $t$ go to zero, which leads to the {\em necessary} condition:  
\bea
\label{eq:BASE_necessary_k}
& & k_{\rm min} \leq k \leq  \bar k + n ,
\nnl 
k_{\rm min}  &= & {\rm max} \left (0, {-h_1 + h_2 - h_3 + h_4 \over 2} \right ) ,
\nnl 
\bar k  &= &  {-3 h_1 + h_2 - h_3 - h_4 \over 2} .
\eea 
For a given $n$ it selects a finite (or empty) set of possible choices of $k$.
For the allowed range of $k$ to be non-empty,  $n$ is bounded by 
\beq
\label{eq:BASE_necessary_n}
n \geq n_{\rm min}, \quad n_{\rm min} =  {\rm max} \left (0, h_1 + h_4, -\bar k\right )  . 
\eeq 
I stress that \eref{BASE_necessary_k} and \eref{BASE_necessary_n} are merely necessary conditions. 
The sufficient condition for $O_{n,k}^{h_1 h_2h_3h_4}$ to be  a local contact term
is that it can be written in the form of \eref{contact}.  
This requires the existence of a solution to the equations relating $h_i$, $n$ and $k$ to the non-negative, integer exponents $n_{ij}$ and $\tilde n_{ij}$. 
That is often more restrictive, thus the allowed range of $k,n$ can be smaller (but never larger) than the one suggested by the necessary conditions. 
In the EFT approach one is usually interested in lowest dimension terms in the $1/\Lambda$ expansion. 
Then it suffices to inspect locality of $O_{n,k}^{h_1 h_2 h_3 h_4}$  for a few values of $n$ close to $n_{\rm min} $ and for the corresponding range of $k$.  

The formulas in Eqs.~(\ref{eq:BASE_twistorcandidate})-(\ref{eq:BASE_necessary_n}) do not treat all incoming particles in the same way. 
This is because arbitrary choices that break the interchange symmetry have been made along the way: 
gauge-fixing $\mu_1$ and $\mu_4$, and eliminating $\langle 1 4 \rangle$ via the Schouten identity.
One could of course alter these choices to arrive at a different family of  basis candidates. 
Note that \eref{BASE_twistorcandidate} is not manifestly parity-invariant because the momentum twistor formalism is not.  
Furthermore, the scaling $\langle 23 \rangle \sim \sqrt u$ implies  that   
\eref{BASE_twistorcandidate} has  a singularity in the $u$-variable for $h_1 - h_2 - h_3 + h_4  < 0$. 
For such helicity configurations  a local term may be a linear combination of $O_{n,k}^{h_1 h_2h_3h_4}$ with different $k$. 
This may be  cumbersome in practice, therefore it is easier  to work with the configurations satisfying $h_1 - h_2 - h_3 + h_4 \geq 0$, and  obtain basis elements for $h_1 - h_2 - h_3 + h_4  < 0$ via the parity operation $P$ acting as $h_i \to - h_i$,  $\lambda_i \leftrightarrow \tilde \lambda_i$.  
The final comment is that \eref{BASE_twistorcandidate} is not automatically symmetric under permutations of external particles $i$, $j$,  even when $h_i =  h_j$. 
For identical particles, symmetrization or anti-symmetrization of the basis elements has to be performed a posteriori.  

To summarize the algorithm, 
the candidate basis elements to span the contact terms  of massless 4-point amplitudes are given in \eref{BASE_twistorcandidate}. 
They are parametrized by two integers: $n$ and $k$.  
The former is constrained by the inequality in \eref{BASE_necessary_n}  and controls the EFT expansion,
 with increasing $n$ corresponding to increasing canonical dimensions. 
For $n$ fixed, $k$ is constrained to a finite range by \eref{BASE_necessary_k}. 
The fact that momentum twistors are unconstrained variables on the manifold of 4-body kinematics ensures the independence of  $O_{n,k}^{h_1 h_2h_3h_4}$ for different $n$ and $k$.  
The elements of this set that are local, meaning they can be written as in \eref{contact}, 
form a basis of contact terms for a given helicity configuration $h_{1,2,3,4}$ and canonical dimension 
$D = 2 n - \sum_i h_i + 4$. 
This prescription is a purely algebraic algorithm to write down a basis of 4-point contact terms in any massless theory. 
 
For practical applications it is more convenient to trade momentum twistors for the standard helicity spinors using 
the identity $(\tilde \mu_2 \tilde \mu_3)  = -  s  \langle 2 3 \rangle  =  \langle 2 3 \rangle^2 [23] $, which follows directly from \eref{MT_defmu}.  
Furthermore, one can simplify spinor expressions using  
$t \langle 12 \rangle \langle 34 \rangle = - s  \langle 13 \rangle \langle 24 \rangle$.  
All in all, \eref{BASE_twistorcandidate} can be recast as 
\bea
\label{eq:BASE_candidate}
O_{n,k}^{h_1 h_2h_3h_4} & = & s^{n-k} t^{k}
\langle 12 \rangle^{-2 h_1} 
 \langle 23 \rangle^{h_1 - h_2 - h_3 + h_4}   
 \nnl & \times & 
   \langle 24 \rangle^{h_1 - h_2 + h_3 - h_4} 
  \langle 34 \rangle ^{-h_1 + h_2 - h_3 - h_4}  . \qquad    
\eea 
This is the central result of this paper.

\subsection{Higher-point}

A similar algorithm as in \sref{4p} can be worked out for higher $N$. 
Any $N$-point contact term can be recast as a sum of rational function of Lorentz-invariant  contractions of 
 $N$ holomorphic spinors $\lambda_1 \dots \lambda_N$ and $N-2$ anti-holomorphic spinors $\tilde \mu_2 \dots \tilde \mu_{N-1}$: 
\beq
O =  \Pi_{k< l = 2}^{N-1} (\tilde \mu_k \tilde \mu_l)^{a_{kl} } 
\Pi_{ i< j = 1}^{N}   (\lambda_i \lambda_j)^{b_{ij} },  
\eeq 
where $a_{kl}$ and $b_{ij}$ are (possibly negative) integers. 
Subsequently, one should mod out dependent contractions using the Schouten identities, 
and relate the exponents $a_{ij}$, $b_{ij}$ to the external helicities. 
Finally, among the irreducible rational functions one should identify the contact terms that can be written in the form of \eref{contact}. 
The procedure is relatively straightforward for small enough $N$, however selection of independent Schouten identities, testing for locality,  and eventual symmetrization for identical particles becomes more pesky with increasing $N$. 
I leave for future publications the details of this procedure and results for concrete physical theories.

\section{Examples}
\label{sec:ex}

This section provides some applications of the master formula  \eref{BASE_candidate} to construct bases of 4-point contact terms in concrete physical theories.  

\subsection{Scalar}

I start with a trivial example of scattering of 4 distinct scalars, $h_i = 0$.  \eref{BASE_candidate} reduces to  
\beq
O_{n,k}= s^{n-k}t^{k} ,   
\eeq 
where $n \geq 0$ from  \eref{BASE_necessary_n}, and $0 \leq k \leq n$ from \eref{BASE_necessary_k}. 
In this case  $O_{n,k}$ simply generates all independent kinematic invariants at a given order in the EFT expansion.  
For $n=0$ ($D$=4, or $\cO(\Lambda^0)$) the only option is $k=0$, that is a constant contact term $O_{0,0} = 1$. 
For $n=1$  ($D$=6 or $\cO(\Lambda^{-2})$) there are 2 options: $k=0$ and $k=1$,  which correspond to the 2 independent Mandelstam invariants $O_{1,0} = s$ and $O_{1,1} = t$. 
For $n=2$ ($\cO(\Lambda^{-4})$)  there are 3 independent invariants 
$O_{2,0}=s^2$, $O_{2,1}= st$, and $O_{2,2} = t^2$.  
And so on... 
For 4 identical scalars one needs to construct linear combinations of the basis elements that are invariant under the $S_4$ permutation symmetry. $O_{0,0}$ is trivially invariant. For $n=1$ no invariant combination exists. 
For $n=2$ the unique permutation-invariant combination is  $O_{2,0}  + O_{2,1} + O_{2,2} = (s^2+t^2+u^2)/2$. 
And so on...

\subsection{Scalar and Spin-h}

A bit less trivial exercise is to determine 4-point contact terms for 2 scalars and 2 identical spin-h particles.
I take $h_{1,4} = 0$ and $h_{2,3} = \pm h$. 
For the both-minus helicity configuration,  \eref{BASE_necessary_k} and  \eref{BASE_necessary_n}
imply $n \geq 0$ and $0 \leq k \leq n$. 
I am interested in the lowest possible $n$, corresponding to the lowest canonical dimension  via \eref{BASE_dim}.
For $n=0$ and $k=0$  \eref{BASE_candidate} yields:  
 \beq 
 \label{eq:EX_omm}
 O^{--} \equiv  O^{--}_{0,0} =  \langle 2 3 \rangle ^{2 h} ,  
  \eeq 
which is manifestly local. 
Thus the lowest order contact term in this helicity sector is unique and is 
$\cO(\Lambda^{-2h})$.
The mirror term in the both-plus sector can be immediately obtained via the parity transformation: 
 \beq
  \label{eq:EX_opp}
 O^{++} \equiv  P \ccdot O^{--} =    [2 3]^{2 h} . 
 \eeq  
 Obtaining the same result directly from \eref{BASE_candidate}  is more tricky, as  $O^{++} $ is a linear combination $\sum_k a_k O^{++}_{2h,k}$ with the coefficients $a_k$ chosen such that the singularity in the $u$ variable cancels out.
This illustrates the fact that the method is not manifestly parity invariant. 
Note that, for any integer or half-integer spin $h$, $O^{--}$ and $O^{++}$ automatically have the correct (anti-)symmetry properties under the exchange $2 \leftrightarrow 3$.  

In the opposite helicity sector, the necessary conditions \eref{BASE_necessary_k} and \eref{BASE_necessary_n} read  $0 \leq k \leq n -h $ and $n \geq h$
This may suggest that the lowest order contact term corresponds to $n = h$. 
However, upon inspection of \eref{BASE_candidate} for $h_2 =  - h_3 = -h$, 
\beq 
O_{n,k}^{-+}  = s^n 
\left ( {\langle 13 \rangle \langle 24 \rangle  \over  \langle 12 \rangle \langle 34 \rangle }  \right )^k 
   \langle 24 \rangle^{2 h } 
  \langle 34 \rangle^{-2h} , 
 \eeq 
it is clear that at least $n=2 h$ is needed\footnote{%
The same conclusion can be reached by using the formula  in Ref.~\cite{Durieux:2019siw} for the minimal dimension of a contact term for a given helicity configuration.}
 to cancel the singularity in $ \langle 34 \rangle$.
This example illustrates the fact that \eref{BASE_necessary_k} and \eref{BASE_necessary_n} are necessary but not sufficient conditions. 
In the end, the lowest order contact term in the $-+$ sector is  $\cO(\Lambda^{-4h})$: 
\beq
\label{eq:EX_omp}
O^{-+} \equiv O^{-+}_{2h,0} =   
\left ( s \langle 2 4 \rangle \over    \langle 34 \rangle \right )^{2h}
= (\lambda_2 p_4 \sigma \tilde \lambda_3)^{2 h}.  
 \eeq 
This is higher order in the EFT expansion than $O^{--}$ and $ O^{++}$ for any $h > 0$.  
The parity mirror is  
$O^{+-} = P \ccdot O^{-+} =  (\lambda_3 p_4 \sigma \tilde \lambda_2 )^{2 h}$.  
 
In summary, at the leading order in the EFT expansion a basis of contact terms for interactions of 2 scalar and 2 spin-h particles is 2-dimensional, consisting of $O^{--}$ and $ O^{++}$. 
These correspond to operators of canonical dimension $D = 2 h + 4$. 
In the language of Lagrangians, the case $h=1/2$ translates to the dimension-5 operators $\phi^2 ( c \psi \psi + \bar c \bar \psi \bar \psi)$, $h = 1$ to   the dimension-6 operators $\phi^2 ( c F_{\mu \nu}  F^{\mu \nu}  + \tilde c  F_{\mu \nu}  \tilde F^{\mu \nu} )$,  
and $h = 2$ to the dimension-8 operators $\phi^2 ( c C_{\mu \nu \alpha \beta}  C^{\mu \nu \alpha \beta}  + \tilde c  C_{\mu \nu \alpha \beta}  \tilde C^{\mu \nu \alpha \beta} )$ where $C_{\mu \nu \alpha \beta}$ is the Weyl tensor~\cite{Ruhdorfer:2019qmk}.  
The expressions in Eqs.~(\ref{eq:EX_omm})-(\ref{eq:EX_omp}) can also be used for $h>2$. 
Of course, massless particles with spin higher than two cannot be consistently coupled to gravity~\cite{Benincasa:2007xk,McGady:2013sga}, 
thus for strictly massless particles this only makes sense as an academic exercise in the limit $\mpl \to \infty$. 
Nevertheless, the structure of the contact terms in the massless limit may give some guidance for constructing a basis of contact terms for {\em massive} higher-spin particles, for which consistent theories exist as EFTs.

\subsection{Euler-Heisenberg}
\label{sec:eh}

The final example is derivation of the leading order contact terms for spin-1 particles, $h_i = \pm 1$. 
I first discuss an academic theory of 4 distinguishable massless spin-1 particles, 
and then restrict to 4 identical particles, aka photons. 
The latter case corresponds to the Euler-Heisenberg Lagrangian.  

Starting with all-minus helicities,  
\eref{BASE_necessary_k}  and \eref{BASE_necessary_n} reduce to $0 \leq k \leq 2+ n$, and $n \geq 0$. 
At the leading EFT order, $n=0$ or $\cO(\Lambda^{-4})$, 
there are 3 independent contact terms corresponding to $k = 0,1,2$: 
\bea
\label{eq:EX_basisspin1mmmm} 
O_1^{----} & =&  {\langle 1 3 \rangle^2  \langle 2 4 \rangle^2},  
\nnl  
O_2^{----} & =&   {s \over t}    {\langle 1 3 \rangle^2  \langle 24 \rangle^2} 
= -{\langle 1 2 \rangle  \langle 1 3 \rangle  \langle 2 4 \rangle \langle 3 4 \rangle} , 
\nnl
O_3^{----} & =&   {s^2  \over t^2 }   \langle 1 3 \rangle ^2  \langle 2 4 \rangle^2  
=  { \langle 1 2 \rangle^2  \langle 34 \rangle^2} .
\eea
where $O_i^{----} \equiv O_{0,i-1}^{----}$.
These are all manifestly local, thus $O_{1,2,3}^{----}$ span a  basis of contact terms in the all-minus sector. 
Exactly the same basis would be obtained via the harmonics method of Ref.~\cite{Henning:2019enq}.   
A basis in the all-plus sector is most easily obtained by the parity operation, 
$O_i^{++++} = P \ccdot O_i^{----}$: 
\bea
\label{eq:EX_basisspin1pppp}
O_{1}^{++++}   & =&  [ 1 3 ]^2  [2 4]^2  ,  \quad 
O_{2}^{++++}  =  -  [ 1 2 ] [ 1 3]  [2 4 ] [3 4 ] , 
\nnl
O_{3}^{++++}  &= &   [1 2]^2  [34]^2.   
\eea 
Equivalently, these elements could be obtained from  \eref{BASE_candidate} for $n=4$ and $k = 0,1,2$. 

Next, consider the configuration $h_{2,3}= -1$, $h_{1,4} = +1$, which implies $n \geq 2$,  $0 \leq k \leq n-2$. 
The leading contact terms corresponds to $n=2$, $k=0$:   
\beq 
\label{eq:EX_m4spin1pmmp}
O^{+--+} \equiv O^{+--+}_{2,0} = s^2 {\langle 2 3 \rangle^4  \over   \langle 1 2 \rangle^2   \langle 3 4 \rangle^2}  
 = \langle 2 3 \rangle^2 [14]^2,  
\eeq
and is also $\cO(\Lambda^{-4})$. 
For the other 2-plus-2-minus configurations the leading contact terms can be trivially obtained by permutations: 
$O^{--++} =  \langle 12 \rangle^2 [34]^2$, 
$O^{-+-+} = \langle 1 3 \rangle^2 [24]^2$, etc. 

For configurations with a single  minus or a single plus helicity \eref{BASE_candidate} does not generate any 
$\cO(\Lambda^{-4})$ local contact terms (they first appear at $\cO(\Lambda^{-6})$). 
All in all, for amplitudes with 4 distinct spin-1 particles, all possible contact  terms are spanned by the following 12 basis elements: 
\beq
\label{eq:EX_gehbasis}
O_{1,2,3}^{----}, \quad O_{1,2,3}^{++++}, \quad O^{--++} + {\rm permutations}. 
\eeq 
In the Lagrangian parlance the the corresponding object is a  basis of dimension-8 operators, e.g.
$F_{\mu \nu}^1 F_{\mu \nu}^2 F_{\alpha \beta}^3 F_{\alpha \beta}^4$,  
$F_{\mu \nu}^1 F_{\mu \nu}^3 F_{\alpha \beta}^2 F_{\alpha \beta}^4$, 
$F_{\mu \nu}^1 F_{\mu \nu}^4 F_{\alpha \beta}^2 F_{\alpha \beta}^3$, 
and 9 analogous one with one or two field strengths replaced by the dual field strength: $F \to \tilde F$.  
One could easily continue this exercise into higher orders in the EFT expansion, simply by incrementing $n$ in the master formula \eref{BASE_candidate}. 
Then \eref{BASE_necessary_k} suggests that, for every helicity configuration, the number of basis elements increases by one at each consecutive EFT order.  

For photons, a basis of contact terms can be constructed as  linear combinations of the elements in \eref{EX_gehbasis} that are invariant under permutations of identical particles.  
In the all-minus and all-plus sectors these are 
\bea
O^{-}  &= &   O_1^{----}  +  O_2^{----} +  O_3^{----} , \nnl  
O^{+} &= & O_1^{++++} + O_2^{++++} +  O_3^{++++} . 
\eea
The four-photon helicity amplitudes are given by 
$A^{++++} = C_+ O^+ + \cO(\Lambda^{-6})$, $A^{----} = C_- O^-  + \cO(\Lambda^{-6})$, 
$A^{--++} = C_0 O^{--++}  + \cO(\Lambda^{-6})$. 
At the leading order in the EFT they are characterized by 3-independent parameters:  
$C_-$, $C_+$, and $C_0$. 
These correspond to the 3 independent dimension-8 four-photon operators : 
\beq
\cL = {1 \over \Lambda^4} \left [ 
a F_{\mu \nu}^2 F_{\alpha \beta}^2
+ b F_{\mu \nu} \tilde F_{\mu \nu} F_{\alpha \beta} \tilde F_{\alpha \beta}
+ c F_{\mu \nu}^2 F_{\alpha \beta} \tilde F_{\alpha \beta} \right ] , 
\eeq 
where the map is  $C_- = 8 (a - b + i c)$,
$C_+ = 8 (a - b - i c)$,
$C_0 = 8 (a + b)$. 
If parity is conserved (as in QED) then $C_- = C_+$, or equivalently $c = 0$, reducing the number of parameters to two.

\section{Summary and Discussion}
\label{sec:disc}

This paper brings you an algorithm for constructing a basis of contact terms of 4-point amplitudes in massless EFTs. 
The master formula is \eref{BASE_candidate}.  
It gives a compact expression for candidate basis elements for any helicity configuration $h_{1,2,3,4}$ of the external particles. 
The candidates $O_{n,k}$  are labeled by two integers $n$ and $k$. 
The former fixes the canonical dimension of the contact term, thus its order in  the EFT expansion. 
At any fixed order there is a finite number  of candidates counted by $k$.
All $O_{n,k}$ in \eref{BASE_candidate} are independent,  but they are not guaranteed to be local. 
Selecting local expressions from this set yields a basis of 4-point contact terms.  
Formally, one can test for locality by searching for solutions to a set of linear equations relating $h_i$, $n$, and $k$ to the exponents $n_{ij}$, $\tilde n_{ij}$ in \eref{contact}, which must be non-negative integers. 
In the 4-point case I find it more practical to pursue a less systematic approach, where first a finite range of allowed $n$ and $k$ is identified by general arguments, for which locality of $O_{n,k}$ can then be quickly assessed by eye. 

It is worth noting that this method directly {\em constructs} basis elements, although it does not a priori {\em count} them.   
In this sense it is orthogonal to the Hilbert series method. 
Indeed, the two can be used together, with the Hilbert series method providing a useful guidance about the EFT order where local contact terms appear and the number of them. 
I verified that for all examples studied in this paper the number of local contact terms generated by   \eref{BASE_candidate} agrees with that predicted by the Hilbert series method~\cite{Brian}.  

It is straightforward to generalize this method to higher-point contact terms, 
but I leave the details to a future publication. 
Less straightforward is to imagine generalization to  EFTs in different spacetime dimensions, or with massive particles. 
Indeed, momentum twistors are only defined for massless particles in four dimensions. 
New clever variables to encode kinematic data seem necessary to perform a similar program beyond 4-dimensional massless EFTs.  

The final comment is that \eref{BASE_candidate} contains more information than just contact terms. 
In particular, it also generates all independent terms with a single pole in the Mandelstam variables.  
This may provide a shortcut to constructing full tree-level amplitudes (rather than just contact terms), which may be useful  especially for  higher-point amplitudes.


\begin{acknowledgments}

I would like to thank  Brian Henning, Tom Melia and Igor Prlina for useful discussions.  
AF is partially supported by the European Union's Horizon 2020 research and innovation programme under the Marie Sk\l{}odowska-Curie grant agreements No 690575 and No 674896.

\end{acknowledgments}

\appendix

\section{GR EFT}
\label{app:gr}

In this appendix I consider the EFT of a massless spin-2 particle, which I call  GR EFT. 
At the lowest order, the theory is the same as the ordinary GR described by the Einstein-Hilbert Lagrangian: 
$\cL_{\rm EH} =-  {\mpl^2 \over 2} \sqrt{-g} R$. 
The higher orders correspond to other coordinate invariant operators added to the Einstein-Hilbert Lagrangian. 
Much as in the rest of this paper, I work at the amplitude level, and only mention Lagrangians for the sake of reference to earlier literature.   
I will start by writing down the most general 3-point amplitudes for spin-2 particles. 
Then I will use \eref{BASE_candidate} to derive the leading contact terms in the 4-point amplitudes for different helicity configurations. 
Finally I will derive the pole terms in the 4-point amplitude required by unitarity, and compare their contributions with that of the contact terms.  

Before beginning, mind that the power counting  here will be somewhat different than in the previous examples.  
Two scales are introduced: the Planck scale $\mpl = (8 \pi G)^{-1/2}$ and the EFT scale $\Lambda$ which is unknown.  
The 3-graviton amplitudes are $\cO(1/\mpl)$, the 4-graviton amplitudes are  $\cO(1/\mpl^2)$, and so on.  
Additional mass scales needed to arrive at a correct dimension of the amplitude will be filled by $\Lambda$.
This way the EFT deformation of GR is more transparent, and the GR limit is recovered for  $\Lambda \to \infty$.  
Of course, one can always identify $\Lambda = \mpl$ a posteriori,  so as to simplify the power counting.  

Start with 3-point amplitudes, the minimal self-interaction of 3 gravitons is described by  
\beq
\label{eq:EGR_Mhhh}
A^{--+} =  -  {\langle 12 \rangle^6 \over  \mpl \langle 13 \rangle^2 \langle 23 \rangle^2 }, 
\quad 
A^{++-} =   -  {[12]^6 \over  \mpl [13 ]^2 [23]^2 }.  
\eeq
This corresponds to the usual  2-derivative graviton cubic interaction in  the Einstein-Hilbert Lagrangian. 
In addition, Poincar\'e symmetry allows for the same-helicity amplitudes: 
\beq
\label{eq:EGR_Mhhh2}
A^{---} =  c_- {\langle 12 \rangle^2 \langle 23 \rangle^2 \langle 31 \rangle^2 \over \mpl \Lambda^4}, \quad
A^{+++}=  c_+ {[12]^2 [23]^2 [31]^2 \over \mpl \Lambda^4}. 
\eeq 
They map to the cubic Weyl tensor terms  in the effective Lagrangian of Ref.~\cite{Ruhdorfer:2019qmk}.  
 
4-point contact terms,  first appear at $\cO(\mpl^{-2} \Lambda^{-6})$ in the EFT expansion.  
The derivation is similar  to the spin-1 case  in \sref{eh}.
In the all-minus sector, for distinguishable spin-2 particles one finds  5 local contact terms at the leading order: 
\beq
O_k^{----}   = 
 (-1)^k \langle 1 3 \rangle^{k}   \langle 24 \rangle^{k}    
  \langle 1 2 \rangle^{4-k}     \langle 3 4 \rangle^{4-k} , 
\eeq 
corresponding to $n=0$ and $k = 0 \dots 4$ in \eref{BASE_candidate}.  
The Bose-symmetric combination is 
\beq
\label{eq:EGR_Oallminus}
O^- \! \equiv   2 O_0^{----} \!+ 2 O_4^{----} \! +  4 O_1^{----} \! + 4 O_3^{----} \!+  6  O_2^{----} , 
\eeq 
which can be simplified as 
\beq
O^- =  \langle 1 2 \rangle^{4}   \langle 34 \rangle^{4}    +   \langle 1 3 \rangle^{4}     \langle 2 4 \rangle^{4} + 
  \langle 1 4 \rangle^{4}     \langle 2 3 \rangle^{4}. 
\eeq 
The  parity mirror of \eref{EGR_Oallminus}, $O^+ = P \ccdot O^-$,  corresponds to replacing $\langle \cdot \rangle$ with 
$[\cdot ]$. 
Moving to the zero total helicity sector, one finds  
\beq
\label{eq:EGR_Ommpp}
O^0 \equiv O^{--++}  = 
\langle 1 2 \rangle^{4}  [34 ]^4 . 
\eeq  
corresponding to $n=4$ and $k=0$  in \eref{BASE_candidate}. 
This one  already has the correct Bose symmetry.  
The 3 contact terms $O^-$, $O^+$ and $O^0$ form a basis at $\cO(\Lambda^{-6})$, and can be mapped to the quartic Weyl tensor operators in the effective Lagrangian of Ref.~\cite{Ruhdorfer:2019qmk}.

Given the 3-point amplitudes and the 4-point contact terms one can construct tree-level 4-graviton amplitudes in GR EFT. 
For each helicity amplitude, the residues on the kinematic poles are calculated by glueing the 3-point amplitudes in \eref{EGR_Mhhh} and \eref{EGR_Mhhh2}.
For the zero total helicity amplitude one finds for $R_s \equiv {\rm Res}_{s\to 0 } \cM^{--++}$:
\bea
R_s & = & -A(1^- 2^- \hat p_s^-) A(3^+ 4^+  p_s^-)  - A(1^- 2^- \hat p_s^+) A(3^+ 4^+  p_s^+) 
\nnl 
&  = &  {\langle 1 2 \rangle^4 [3 4 ]^4 \over \mpl^2} \left [ {1 \over t u }  + c_- c_+  { t u \over \Lambda^8 }  \right ] .  
 \eea   
 From there the full amplitude can be reconstructed as 
 \beq
 \label{eq:EGR_Ammpp}
 A^{--++} = {\langle 1 2 \rangle^4 [3 4 ]^4 \over \mpl^2}  \left [ {1 \over s t u}  + {d_0 \over \Lambda^6}  + \cO(\Lambda^{-8}) \right ]  . 
 \eeq 
The first term in the square bracket yields the well known result for tree-level graviton scattering in GR. 
One can verify that the residues of \eref{EGR_Ammpp} in the $t$ and $u$ channels are also consistent with the factorization into 3-point amplitudes.  
The leading  EFT correction comes from the contact term in  \eref{EGR_Ommpp}.
The pole contribution proportional to $c_- c_+$ is neglected above because it is subleading. 
To include it consistently one should also calculate $\cO(\Lambda^{-8})$ contact terms 
(which can be done easily starting from \eref{BASE_candidate}, if ever needed). 
In the all-minus and all-plus sector the discussion is simpler, as only the contact terms contribute at tree level: 
\bea
 \label{eq:EGR_Ammmm}
 A^{----} &= & {d_- \over \mpl^2 \Lambda^6} \bigg ( 
 \langle 1 2 \rangle^{4}   \langle 34 \rangle^{4}    +   \langle 1 3 \rangle^{4}     \langle 2 4 \rangle^{4} + 
  \langle 1 4 \rangle^{4}     \langle 2 3 \rangle^{4} \bigg ), 
  \nnl
 A^{++++} &= & {d_+ \over \mpl^2 \Lambda^6} \bigg (
 [ 1 2 ]^{4}  [34]^{4}    +  [1 3 ]^{4}    [ 2 4 ]^{4} + 
[1 4 ]^{4}  [ 2 3]^{4} \bigg ), 
\nnl
\eea
up to $\cO(\Lambda^{-8})$ corrections.   
The EFT corrections interfere with GR in  \eref{EGR_Ammpp} but not in \eref{EGR_Ammmm}. 
Therefore the leading EFT effect on graviton scattering observables (proportional to  the helicity amplitudes squared) is determined by the contact term $O^{--++}$ and parametrized by a single Wilson coefficient  $d_0$.

\bibliographystyle{apsrev4-1}
\bibliography{twister}

\end{document}